\documentclass[apj]{emulateapj}
\usepackage{mathptmx}

\def\gtorder{\mathrel{\raise.3ex\hbox{$>$}\mkern-14mu
             \lower0.6ex\hbox{$\sim$}}}
\def\ltorder{\mathrel{\raise.3ex\hbox{$<$}\mkern-14mu
             \lower0.6ex\hbox{$\sim$}}}

\slugcomment{Draft of \today}

\shorttitle{SN~2006gy}

\shortauthors{Ofek et al.}

\begin{document}

\title{SN~2006gy: An extremely luminous supernova in the galaxy NGC\,1260}
\author{
E.~O.~Ofek\altaffilmark{1},
P.~B.~Cameron\altaffilmark{1},
M.~M.~Kasliwal\altaffilmark{1},
A.~Gal-Yam\altaffilmark{1},
A.~Rau\altaffilmark{1},
S.~R.~Kulkarni\altaffilmark{1},
D.~A.~Frail\altaffilmark{2},
P.~Chandra\altaffilmark{2}$^{,}$\altaffilmark{3},
S.~B.~Cenko\altaffilmark{1},
A.~M.~Soderberg\altaffilmark{1}
\&\
S.~Immler\altaffilmark{4},
}
\altaffiltext{1}{Division of Physics, Mathematics and Astronomy, California Institute of Technology, Pasadena, CA 91125, USA}
\altaffiltext{2}{National Radio Astronomy Observatory, Charlottesville VA 22903}
\altaffiltext{3}{University of  Virginia, Charlottesville VA 22904}
\altaffiltext{4}{NASA/CRESST/GSFC, Code 662, Greenbelt, MD 20771, USA}

\begin{abstract}

With an extinction-corrected V-band peak absolute magnitude of
about $-22$, supernova (SN)~2006gy is probably the brightest SN
ever observed.  
We report on multi-wavelength observations of this
SN and its environment.
Our spectroscopy shows
an H$\alpha$ emission line
as well as absorption features which
may be identified as \ion{Si}{2} lines
at low expansion velocity.
The high peak luminosity, the slow rise to maximum,
and the narrow H$\alpha$ line
are similar to those observed in
hybrid type-Ia/IIn (also called IIa) SNe.
The host galaxy, NGC\,1260,
is dominated by an old stellar population with solar metallicity.
However, our high resolution adaptive optics images reveal
a dust lane in this galaxy,
and there appears to be an
\ion{H}{2} region in the vicinity of the SN.
The extra-ordinarily large peak luminosity, 
$\sim 3\times 10^{44}\,$erg\,s$^{-1}$, demands a dense
circum-stellar medium, regardless of the mass of the
progenitor star. The inferred mass loss rate of the
progenitor is
$\sim0.1\,$M$_{\odot}\,$yr$^{-1}$ 
over a period of $\sim10\,$yr prior to explosion.
Such an high mass-loss rate may be the result
of a binary star common envelope ejection.
The total radiated energy in the first two months
is about $1.1\times 10^{51}\,$erg, which is only a factor of two
less than that available from a super-Chandrasekhar Ia explosion.
Therefore, given the presence of a star forming region in
the vicinity of the SN and the high energy requirements,
a plausible scenario is that SN~2006gy
is related to the death of a massive star
(e.g., pair production SN).

\end{abstract}


\section{Introduction}
\label{Introduction}


SN\,2006gy
was discovered by the ROTSE-IIIb telescope at
the McDonald Observatory on UT 2006 September 18.3 (Quimby 2006).
The supernova (SN) was initially reported
$2''$ off the center of NGC\,1260.
Harutyunyan et al. (2006) obtained a spectrum on
UT 2006 September 26 and reported 
a three-component H$\alpha$ emission line:
an unresolved narrow line; an intermediate component
with
Full Width at Half Maximum (FWHM) of 
$2500\,$km\,s$^{-1}$; and a component with FWHM of $9500\,$km\,s$^{-1}$.
They suggested that the event was a type~II SN.

Prieto et al. (2006)
reported that a spectrum of the SN,
obtained eight days after the discovery,
was suggestive
of a dust-extinguished type-IIn event.
However, the Balmer lines were symmetric, which is unusual
for SNe in their early phases.
Moreover, after correcting for two magnitudes of extinction
(based on observed Na\,I lines), the absolute magnitude
is about $-22$.
They further reported that the position of the SN is consistent
with the center of the galaxy, suggesting
it is more consistent with
an eruption of the active galactic nucleus (AGN) of 
NGC\,1260.
Foley et al. (2006) noted that the SN is offset by about $1''$ from
the nucleus of NGC\,1260.
This fact along with a spectrum obtained six days after discovery,
led them to suggest that SN\,2006gy was a type IIn event.

Here we report on
multi-wavelength observations of SN\,2006gy.
An independent contemporary analysis is presented
by Smith et al. (2006).

\section{Observations}
\label{Obs}

\begin{figure}
\centerline{\includegraphics[width=8.5cm]{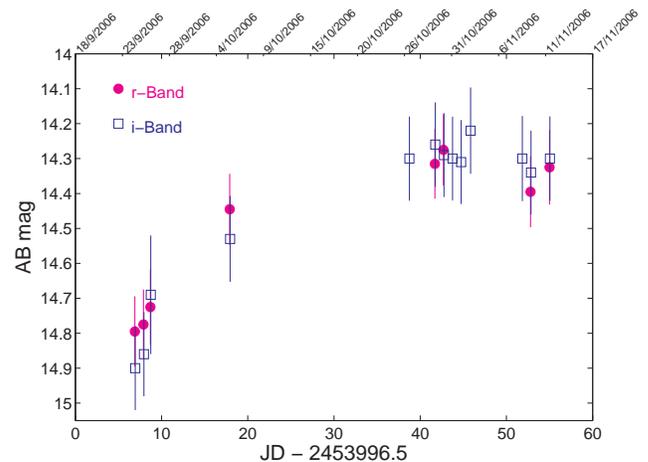}}
\caption{$r$-band (filled circles) and $i$-band (empty squares)
light curves
of SN\,2006gy. 
Errors include the uncertainty in the absolute
calibration. The bottom horizontal axis
shows the time relative to the discovery
of SN\,2006gy (2006 September 18).
\label{SN2006gy_LC}}
\end{figure}

We initiated a photometric ($g,r,i,z$) monitoring program with the 
Palomar 60-inch robotic telescope (Cenko et al. 2006).
Not possessing pre-explosion images of NGC\,1260, which are essential
for accurate subtraction of the light from the host galaxy, we used
archival $R$- and $I$-band images obtained with the 
Jacobus Kapteyn Telescope\footnote{ING archive: http://casu.ast.cam.ac.uk/casuadc/archives/ingarch}
on 1996 January 13 and 1991 December 1, respectively.
The $r$- and $i$-band measurements and errors, presented 
in Fig.~\ref{SN2006gy_LC} and Table~\ref{Tab-Log}
were produced by image subtraction using the
Common Point-spread-function Method
(CPM; Gal-Yam et al. 2004).

\begin{deluxetable}{lll|lll}
\tablecolumns{6}
\tablewidth{0pt}
\tablecaption{Log of observations and measurements}
\tablehead{
\colhead{Date} &
\colhead{Tel} &
\colhead{Magnitude\tablenotemark{a}} &
\colhead{Date} &
\colhead{Tel} &
\colhead{Magnitude} \\
\colhead{UT 2006} &
\colhead{\& Band} &
\colhead{}        &
\colhead{UT 2006} &
\colhead{\& Band} &
\colhead{or Flux\tablenotemark{a}}
}
\startdata
09-24.9 & P60\tablenotemark{b} $r$ & $14.80\pm0.10$      & 10-30.7 & P60 $i$ & $14.29\pm0.12$\\
09-25.9 &         & $14.78\pm0.10$                       & 10-31.7 &         & $14.30\pm0.12$\\
09-26.7 &         & $14.73\pm0.11$                       & 11-01.7 &         & $14.31\pm0.12$\\
10-05.9 &         & $14.45\pm0.10$                       & 11-02.8 &         & $14.22\pm0.12$\\
10-29.7 &         & $14.32\pm0.10$                       & 11-08.8 &         & $14.30\pm0.12$\\
10-30.7 &         & $14.28\pm0.10$                       & 11-09.8 &         & $14.34\pm0.12$\\
11-09.8 &         & $14.40\pm0.10$                       & 11-12.0 &         & $14.30\pm0.12$\\
11-12.0 &         & $14.33\pm0.11$                       & 11-01.3 &P200\tablenotemark{c} $J$ & $12.96\pm0.14$\\
09-24.9 & P60 $i$ & $14.90\pm0.12$                       & 11-01.3 &P200 $K_{\rm s}$          & $12.59\pm0.17$\\
09-25.9 &         & $14.86\pm0.12$                       & & & \\
09-26.7 &         & $14.69\pm0.17$                       & & & \\
10-05.9 &         & $14.53\pm0.12$                       & 11-20.4 & VLA X\tablenotemark{d}  & $186\pm80\,\mu$Jy \\
10-26.7 &         & $14.30\pm0.12$                       & 11-20.4 & VLA K\tablenotemark{d}  & $59\pm110\,\mu$Jy \\
10-29.7 &         & $14.26\pm0.12$                       & 11-23.2 & VLA Q\tablenotemark{d}  & $56\pm120\,\mu$Jy \\
\enddata
\tablenotetext{a}{Observed magnitude or flux density
of the SN. Magnitude errors include the uncertainty in absolute calibration, which dominates the errors.
To convert specific-flux errors to $3$-$\sigma$ upper limits multiply the errors by $3$.}
\tablenotetext{b}{Palomar 60-inch (P60) magnitudes are given in the AB magnitude system.
Absolute calibration was performed by fitting the
Hipparcos $B_{T}V_{T}$ and 2MASS (Skrutskie et al. 2006)
$JHK$ magnitudes of three nearby Tycho-2 (H{\o}g et al. 2000)
reference stars to synthetic photometry of
stellar spectral templates (Pickles 1998) in the same bands.
The best fit spectral template of each star was used
to calculate its synthetic magnitudes in the $r$- and $i$-bands.
The uncertainty in this calibration process, calculated from the
scatter between the zero-points derived from each Tycho-2 star,
is about $0.1\,$mag.}
\tablenotetext{c}{Palomar 200-inch (P200) IR Vega-based PSF-fitting magnitudes, relative to the 2MASS star 03172629$+$4124103 within the field, as measured with IRAF/DAOPHOT.}
\tablenotetext{d}{Center frequency of VLA bands are as follows:
8.4\,GHz (X), 22.5\,GHz (K) and 43.3\,GHz (Q).}
\label{Tab-Log}
\end{deluxetable}
%

\begin{figure}
\centerline{\includegraphics[width=8.5cm]{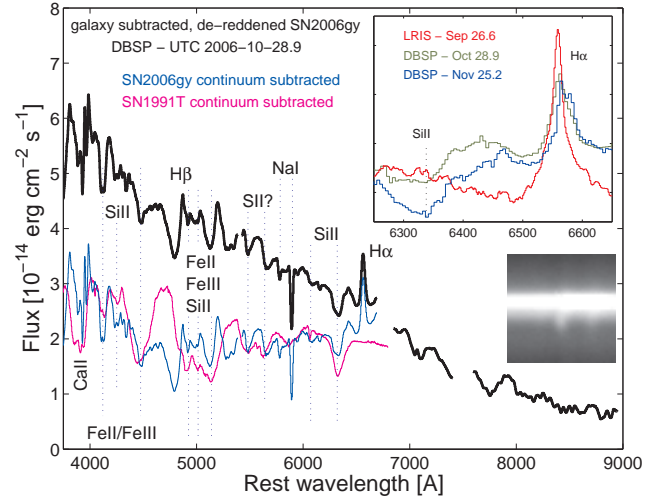}}
\caption{The spectrum of SN\,2006gy after subtracting
the scaled S0 template, and correcting
for Milky-Way and NGC\,1260 extinction (black line; see text).
The blue line shows the same spectrum after subtraction
of a third-degree polynomial fitted to the spectrum.
The magenta line shows the spectrum of the
luminous type-Ia SN\,1991T at nine days post peak brightness,
after the same processing.
The spectrum of SN\,1991T was redshifted by $\sim8500\,$km\,s$^{-1}$
in order that the possible \ion{Si}{2} features in both spectra coincide.
A zoom-in on the H$\alpha$ emission line
as observed by LRIS and DBSP is shown in the upper inset.
The lower inset shows a section of the 2-dimensional Keck
spectrum of SN\,2006gy obtained under good seeing conditions
on 2006 December 18.
The extension in the spatial (vertical) direction is an H$\alpha$ emission
near the SN location
(as previously reported by Smith et al. 2006).
{\bf Technical details}:
For the LRIS spectra,
with integration time ranging from 600 to $2400\,$s,
we employed
the $1\farcs5$ slit with the 400/8500 grating blazed at $7550\,$\AA,
and the 600/4000 grism, in the red and blue sides, respectively.
One exception is the December 19 spectrum that was obtained using the
high-resolution R1200/7500 grating centered on the \ion{Na}{1} lines.
The DBSP observations were obtained with 
a $1\farcs5$ slit and R158/7500
and B600/4000 gratings on the red and blue
arms, respectively. The spectrum marked with October 28.9
is the sum of 
four spectra (total integration time of $1500\,$s)
obtained during the October run.
The integration time for
the November DBSP spectrum was $900\,$s.}
\label{Spec_DBSP_SN}
\end{figure}

On UT 2006 September 26 and December 18 and 19
we obtained spectra using the
Low Resolution Imaging Spectrograph mounted on the Keck-I 10-m
telescope (LRIS; Oke et al. 1995).
Spectra were also obtained
on UT 2006 October 28.3, 29.4 and November 25.2, using the Double
Beam Spectrograph (DBSP) mounted on the Hale 5-m telescope.
The spectra are displayed in Fig.~\ref{Spec_DBSP_SN}.

%

On UT 2006 Nov 1.3 we observed the event with
the Adaptive Optics system (Troy et al. 2000)
equipped with the Palomar High Angular Resolution Observer
(Hayward et al. 2001)
camera mounted on the Hale 5-m telescope.
We used the
wavefront reconstruction algorithm --
denominator-free centroiding and Bayesian reconstruction (Shelton 1997),
which delivered $K_{\rm s}$-band images with
$0\farcs1$ FWHM and a Strehl ratio of $\sim15\%$.
We obtained $660\,$s and $300\,$s images in the
$K_{\rm s}$ and $J$-bands, respectively, using
the high-resolution mode ($25\,$mas\,pix$^{-1}$)
and a $240\,$s $K_{\rm s}$-band image using
the low-resolution camera ($40\,$mas\,pix$^{-1}$).
Each frame was flat-fielded, background
subtracted, and repaired for bad pixels using custom PyRAF
software\footnote{PyRAF is a product of Space Telescope Science
Institute, which is operated by AURA for NASA.}.
\begin{figure}
\centerline{\includegraphics[width=8.5cm]{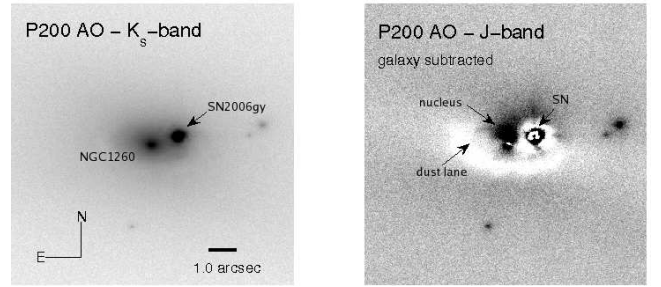}}
\caption{{\bf Left}:
$K_{\rm s}$ band image.
The SN (marked) is clearly resolved
from the galaxy nucleus.  {\bf right}: $J$-band image after subtracting
the best fit S\'{e}rsic profile from the galaxy and a Gaussian
profile from the SN using GalFit (Peng et al. 2002).  
The S\'{e}rsic
model parameters are as follows: index of $3.7$, an effective radius of
$34''$, an axial ratio of $0.51$, and a position angle of $80\,$deg.
A dust lane (white band) is seen southward of the galaxy nucleus.
Based on three 2MASS sources, 
we derived the galaxy nucleus position:
$\alpha=03^{h}17^{m}27.^{s}241$, $\delta=+41^{\circ}24^{m}18\farcs55$ 
and the SN position (end numbers):
$27^{s}.158$ ($\alpha$) and $18\farcs88$ ($\delta$).
The astrometric solution has
rms of $0\farcs04$ and $0\farcs01$ in $\alpha$ and $\delta$, respectively.
\label{P200_AO}}
\end{figure}
%

The field of SN\,2006gy/NGC\,1260 was observed by the {\it Swift}
X-Ray Telescope (XRT) on 2006 October 30 and the {\it Chandra}
X-ray Observatory  on 2004 December 23 and 2006 November 14\footnote{This
latest observation was conducted under Director's discretionary
time (PI: Pooley).}.
For the {\it Swift} observations, 
assuming a Galactic neutral
Hydrogen column density $N_{H}=1.3\times10^{21}\,$cm$^{-2}$
(Dickey \& Lockman 1990),
and a power-law spectrum with index $1.8$,
we set a $3$-$\sigma$ upper limit for the flux in 
the $0.2-10\,$keV band 
of $<1.8\times10^{-13}\,$erg\,s$^{-1}$\,cm$^{-2}$.
The {\it Chandra} observations reveal a variable source
at the position of
the nucleus of NGC\,1260. The spatial coincidence lead us to attribute
this source to an active galactic nucleus.
In order to constrain the X-ray luminosity of the SN,
we fitted the X-ray image with a model containing three components:
A narrow Gaussian centered on the galaxy position;
a wide Gaussian centered on the galaxy position (i.e., diffuse emission);
and a narrow Gaussian centered on the SN position.
We find that the SN flux is consistent with zero,
and that
its flux is $<1.6\times10^{40}\,$erg\,s$^{-1}$
at the $3$-$\sigma$ confidence level,
assuming a photon index of 1.8, a distance of 73\,Mpc to
NGC\,1260 and a neutral Hydrogen column density of
$N_{H}=6.3\times10^{21}\,$cm$^{-2}$ (in the Galaxy and NGC\,1260).
%
This field was also observed using the {\it Swift}
Ultraviolet/Optical Telescope.
We are awaiting
late epoch observations in order to properly remove the host
contamination.

We performed radio observations of NGC\,1260 with the Very Large Array
(VLA)\footnote{The Very Large Array is operated by the National Radio
Astronomy Observatory, a facility of the National Science Foundation
operated under cooperative agreement by Associated Universities, Inc.}
on 2006 Nov 20 and 23 UT. 
The observations were obtained in continuum 
mode with a bandwidth of $2\times50$ MHz. We observed 3C\,48
(J0137$+$331) for flux calibration, while phase referencing was
performed against J0319$+$415. The data were reduced using
the Astronomical Image Processing System.
We did~not detect a source at
the position of the SN (see Table~\ref{Tab-Log}).

\section{Spectral Analysis}
\label{Spec}

%
%
In the optical spectra we identify two \ion{Na}{1}
absorption lines, one of Galactic origin and
the other at the redshift of NGC\,1260 ($z=0.019$).
Based on the ratio between the equivalent widths
of these two absorption lines (e.g., Munari \& Zwitter 1997),
we estimate the total extinction
toward SN\,2006gy to be $\sim4.4$ times
the Galactic extinction ($E_{B-V}=0.16$; Schlegel et al. 1998),
which gives $E_{B-V}\approx0.7$.
The high-resolution spectrum,
obtained on 2006 December 19,
resolves the \ion{Na}{1} doublet. Based on this spectrum, we find that
both the Galactic and NGC\,1260 doublets have similar line ratios
and are not saturated.
We note that using the \ion{Na}{1}-extinction correlation
derived by Turatto et al. (2003), we find
a total $E_{B-V}$ extinction in the range $1$ to $3.5\,$mag.
Moreover, the extinction is derived by
assuming that all the \ion{Na}{1} absorption 
is of light emitted by the SN, rather than of light emitted
by the host galaxy.
Therefore, the extinction toward SN\,2006gy is
uncertain and this issue will require further study.
%
%
%
%

In Fig.~\ref{Spec_DBSP_SN} we show the DBSP spectrum
of SN\,2006gy (black line)
after the subtraction of
a scaled S0 galaxy template (Kinney et al. 1996).
The template was reddened
to account for Galactic extinction in the direction
of the SN, and
scaled so that the
synthetic $r-i$ color
of the host-subtracted SN
spectrum
matches the photometrically observed value at the same epoch.
Next, we flux-calibrated the spectrum by requiring its $r$-band
synthetic photometry to equal
the observed magnitude of the SN at the same epoch.
Finally, we corrected the spectrum for
extinction,
assuming $E_{B-V}\approx0.7\,$mag
and $R_{V}=3.08$ (Cardelli, Clayton, \& Mathis 1989).
From the final spectrum we find, at maximum light, 
an extinction corrected
synthetic V-band magnitude (Vega system) of about $12.4$ mag.

Our spectra show an H$\alpha$ and H$\beta$ emission lines
with a P-cygni profile, characteristic of type-IIn SNe.
%
%
%
%
%
We note that the equivalent width of the H$\alpha$ line
is decreasing with time.
%
Interestingly, we detected several absorption features
which may be
\ion{Si}{2},
\ion{S}{2},
\ion{Fe}{2}, \ion{Fe}{3} and \ion{Ca}{2} lines.
Such lines are usually observed in type-Ia SNe.
%
However, we stress that the
relative lines strenghts and apparent low expansion
velocities (i.e., $\sim 1000$--$2000\,$km\,s$^{-1}$)
are peculiar, making
line identifications {\it tentative only}.
To emphasize this,
we show in Fig.~\ref{Spec_DBSP_SN} the spectrum
of SN\,1991T (Filippenko et al. 1992)
at nine~days from maximum light redshifted by $8500\,$km\,s$^{-1}$,
and the spectrum
of SN\,2006gy at 42 days since discovery, after the subtraction
of third-degree polynomials fitted to each spectrum.
The lines of SN\,2006gy are narrower and red-shifted
relative to these of SN\,1991T,
indicating that SN\,2006gy had a lower expansion velocity
compared to type-Ia SN.

\section{Environment}
\label{Env}

NGC\,1260 is an early-type galaxy within the Perseus cluster of galaxies.
Its Heliocentric recession velocity is
$5760\,$km\,s$^{-1}$ and its velocity dispersion is
$201\pm12\,$km\,s$^{-1}$ (Wegner et al. 2003).
Based on the recession velocity of the cluster 
the distance modulus to NGC\,1260 is
$34.5\,$mag\footnote{Assuming Hubble parameter $H_{0}=72\,$km\,s$^{-1}\,$Mpc$^{-1}$, matter content $\Omega_{m}=0.27$, and dark-energy content of $\Omega_{\Lambda}=0.73$.}.


Our adaptive optics images (Fig.~\ref{P200_AO}) show that
SN\,2006gy is located $0\farcs99$ (projected distance $380\,$pc), at a
position angle of $290\,$deg, from the nucleus of NGC\,1260.
A dust lane,
passing about $300\,$pc (projected)
from the SN location, is clearly seen
in our galaxy-subtracted $J$-band image.
Moreover, we confirm the detection by
Smith et al. (2006) of an \ion{H}{2} region
in the SN vicinity (Fig.~\ref{Spec_DBSP_SN}).

The Mg$_2$ index of this galaxy was
measured to be in the range $0.24$--$0.27\,$mag
(Davis et al. 1987; Wegner et al. 2003).
This value, along with the synthetic
spectral models of Vazdekis (1999),
suggests that the metallicity of NGC\,1260 is
not low, [Fe/H]$\ga-0.2$.

\section{Discussion}
\label{Disc}

With estimated peak absolute magnitude of $V\approx-22$,
SN\,2006gy is probably the brightest SN ever observed.
The
slow brightening, the peak luminosity and the 
H$\alpha$ emission line and the possible
SN-Ia-like features suggest 
that SN\,2006gy maybe related
to the hybrid IIn/Ia SNe class
(also known as type-IIa; Deng et al. 2004).
The other possible members in the type-IIa group are
SN\,2002ic (Hamuy et al. 2003)
SN\,2005gj (Aldering et al. 2006);
SN\,1997cy (Germany et al. 2000)
and SN\,1999E (Rigon et al. 2003).
%
%

Any model of SN\,2006gy has to explain 
the spectral lines,
the extra-ordinary peak luminosity of 
$L_{p}\sim3\times10^{44}\,$erg\,s$^{-1}$ (after correction for
extinction), and a radiated energy over the first two months
of $E_{rad}\sim1.1\times10^{51}\,$erg
(assuming 11,000-K black body which roughly matches the
Rayleigh-Jeans slope in DBSP spectra).
We note that even if the extinction in NGC\,1260 was overestimated
and the SN light suffers only from Galactic extinction, the total
radiated energy within the first two months is about $3\times10^{50}\,$erg.

The high peak luminosity suggests that the blast wave
from the explosion
efficiently converts the mechanical energy to radiation.
This mean the shock has to be radiative
which requires the circum-stellar medium (CSM) density to
exceed $10^6\,{\rm cm}^{-3}$.
Moreover, the conversion of mechanical energy
of an explosion to radiation requires
that the ejecta
sweep up matter with comparable mass.
The slow rise time, $t_{p}\sim50\,$d
to peak luminosity implies that the
dense region has a size of at least $R\sim v_st_p\sim 2\times 10^{15}\,{\rm
cm}$, where $v_s\sim 5\times 10^8\,{\rm cm\,s}^{-1}$ is the speed of the
blast wave.
The peak luminosity, $L_{p}$, requires density
of the order,
$n\sim L_{p}/(2\pi R^{2} v_{s}^3)\sim 10^{10}\,$cm$^{-3}$.
Assuming an upper limit on $v_{s}$ of $10^{9}\,$cm\,s$^{-1}$
at early times,
the minimum mass contained within this radius is $\gtorder0.2\,$M$_\odot$.
The gradual decrease in the radiated
energy, and possibly lower expansion velocities
would easily
bring it closer to a solar mass.
The mass loss
rate by the progenitor has to be stupendous, $\dot M \sim
1\,M_\odot/(t_{p} v_{s}/v_{w})\sim 10^{-1}\,M_\odot\,{\rm yr}^{-1}$,
over a time scale of at least about $10\,$yr,
where $v_{w}=200\,$km\,s$^{-1}$
is the speed of the progenitor wind
(Smith et al. 2006; $v_{w}\sim 130$--$260\,$km\,s$^{-1}$).
Finally, the high CSM density accounts for the lack of
substantial X-ray and radio emission (being absorbed by photoelectric
and free-free absorption, respectively).

SN2006gy shares some properties with type-IIa and type
IIn SNe. Type-IIn SNe are most plausibly the result
of a core collapse SN embedded in dense CSM, while
IIa events have been explained as thermo-nuclear explosions
taking place in a dense medium
(e.g., Livio \& Riess 2003; Han \& Podsiadlowski 2006).
%
%
The thermo-nuclear model is attractive from a spectroscopic
perspective.
In the context of type-Ia SNe,
a possible explanation to the high-mass loss rate
is that it is the
result of a common-envelope phase in a binary system (e.g., Taam
\& Ricker 2006 and references therein).  This scenario was suggested
by Livio \& Riess (2003) to explain the properties of SN\,2002ic,
and is consistent with the inferred high mass loss rate
and its velocity (i.e., $\sim200\,$km\,s$^{-1}$).
However,
this scenario
requires the ejection of matter from the
progenitor to shortly precede the SN explosion
(Chugai \& Yungelson 2004).
Moreover, the
total kinetic energy of Ia events is limited to
about $1$--$2\times10^{51}\,$erg (Khokhlov et al. 1993),
and it can get up to $2.5\times10^{51}\,$erg for
super-Chandrasekhar models (cf. Yoon \& Langer 2004).
Therefore,
unless we considerably over-estimated the extinction,
the total radiated energy of SN\,2006gy
in the first two months alone is challenging for type-Ia-like
SN models.

Smith et al. (2006),
noting that the envelope of a massive star
($>100\,$M$_{\odot}$)
contains a reservoir
of thermal energy that can power the SN,
suggested that such a star was the progenitor of SN\,2006gy.
However, most of the thermal energy
will be lost due to expansion and
the ability of the photons to leak out is
limited by the long diffusion timescale for photons
($\gtorder$months; e.g., Kulkarni 2005).
Therefore, it will be difficult for this specific model
(alone) to explain the high
peak luminosity of SN\,2006gy.

Along the general lines of previous suggestion
by Benetti et al. (2006; for SN\,2002ic)
and Smith et al. (2006; for SN\,2006gy),
we speculate that
the large energy budget for SN\,2006gy
may hint at a highly energetic explosion ($\sim10^{52}\,$erg),
from a massive stellar progenitor.
Two possibilities are an CSM-embedded collapsar
(e.g., Woosley \& MacFadyen 1999),
or a pair production SN
(e.g., Ober et al. 1983; Smith et al. 2006).
Pair production SNe, however, require low-metallicity progenitors,
but it may be possible to overcome this requirement
by the merger of two massive stars.
We further speculate that such a merger may be responsible
to the high mass-loss rate (e.g., common envelope ejection).
SN\,2006gy and other IIa (and maybe many IIn) 
events may result from one of these energetic explosions
that are able to produce $\sim10^{52}\,$erg.

The general issues of the large energy release into a dense CSM
have been discussed for some type IIn events 
(e.g., Chugai et al. 2004, Gal-Yam et al. 2006).
For reasons
we do not understand the explosion is preceded by a phase of
stupendous mass loss.
The mass and geometry of the hydrogen envelope may determines
the outcome of the explosion (e.g., IIa or IIn).
Finally, the rarity of such energetic SNe
reflect the rarity of the progenitors.

\acknowledgments
We are grateful to N. Gehrels for approving the {\it Swift} observations.
We thank Re'em Sari, Sterl Phinney, Orly Gnat, Ehud Nakar and Lauren MacArthur
for valuable discussions, and we are grateful to J. Hickey for
his help obtaining the AO observations, and to D. Sand and R. Ellis
for spectroscopic observations.
This work is supported in part by grants from NSF and NASA.
This research was partially based on data from the ING Archive.


\begin{thebibliography}{}




\bibitem[Aldering et al.(2006)]{2006ApJ...650..510A} Aldering, G., et al.\ 
2006, ApJ, 650, 510 

\bibitem[Benetti et al.(2006)]{2006astro.ph.11125B} Benetti, S., 
Cappellaro, E., Turatto, M., Taubenberger, S., Harutyunyan, A., \& Valenti, 
S.\ 2006, astro-ph/0611125 


\bibitem[Cardelli et al.(1989)]{1989ApJ...345..245C} Cardelli, J.~A., 
Clayton, G.~C., \& Mathis, J.~S.\ 1989, ApJ, 345, 245 

\bibitem[Cenko et al.(2006)]{2006PASP..118.1396C} Cenko, S.~B., et al.\ 
2006, PASP, 118, 1396 

\bibitem[Chugai \& Yungelson(2004)]{2004AstL...30...65C} Chugai, N.~N., \& 
Yungelson, L.~R.\ 2004, Astronomy Letters, 30, 65 

\bibitem[Chugai et al.(2004)]{2004MNRAS.355..627C} Chugai, N.~N., 
Chevalier, R.~A., \& Lundqvist, P.\ 2004, MNRAS, 355, 627 



\bibitem[Deng et al.(2004)]{2004ApJ...605L..37D} Deng, J., et al.\ 2004, 
ApJL, 605, L37 

\bibitem[Dickey \& Lockman(1990)]{1990ARA&A..28..215D} Dickey, J.~M., \& 
Lockman, F.~J.\ 1990, ARA\&A, 28, 215 

\bibitem[Filippenko et al.(1992)]{1992ApJ...384L..15F} Filippenko, A.~V., 
et al.\ 1992, ApJL, 384, L15 

\bibitem[Foley et al.(2006)]{2006CBET..695....1F} Foley, R.~J., Li, W., 
Moore, M., Wong, D.~S., Pooley, D., \& Filippenko, A.~V.\ 2006, Central 
Bureau Electronic Telegrams, 695, 1 (2006).~ Edited by Green, D.~W.~E., 
695, 1 

\bibitem[Gal-Yam et al.(2004)]{2004ApJ...609L..59G} Gal-Yam, A., et al.\ 
2004, ApJL, 609, L59 

\bibitem[Gal-Yam et al.(2006)]{2006astro.ph..8029G} Gal-Yam, A., et al.\ 
2006, ApJ, in press, astro-ph/0608029 

\bibitem[Germany et al.(2000)]{2000ApJ...533..320G} Germany, L.~M., Reiss, 
D.~J., Sadler, E.~M., Schmidt, B.~P., \& Stubbs, C.~W.\ 2000, ApJ, 533, 
320 

\bibitem[Hamuy et al.(2003)]{2003Natur.424..651H} Hamuy, M., et al.\ 2003, 
Nature, 424, 651 

\bibitem[Han \& Podsiadlowski(2006)]{2006MNRAS.368.1095H} Han, Z., \& 
Podsiadlowski, P.\ 2006, MNRAS, 368, 1095 

\bibitem[Harutyunyan et al.(2006)]{2006CBET..647....1H} Harutyunyan, A., 
Benetti, S., Turatto, M., Cappellaro, E., Elias-Rosa, N., \& Andreuzzi, G.\ 
2006, Central Bureau Electronic Telegrams, 647, 1 (2006).~ Edited by Green, 
D.~W.~E., 647, 1 

\bibitem[Hayward et al.(2001)]{2001PASP..113..105H} Hayward, T.~L., Brandl, 
B., Pirger, B., Blacken, C., Gull, G.~E., Schoenwald, J., \& Houck, J.~R.\ 
2001, PASP, 113, 105 

\bibitem[H{\o}g et al.(2000)]{2000A&A...355L..27H} H{\o}g, E., et al.\ 
2000, A\&A, 355, L27 




\bibitem[Khokhlov et al.(1993)]{1993A&A...270..223K} Khokhlov, A., Mueller, 
E., \& Hoeflich, P.\ 1993, A\&A, 270, 223 

\bibitem[Kinney et al.(1996)]{1996ApJ...467...38K} Kinney, A.~L., Calzetti,
D., Bohlin, R.~C., McQuade, K., Storchi-Bergmann, T., \& Schmitt, H.~R.\
1996, ApJ, 467, 38


\bibitem[Kulkarni(2005)]{2005astro.ph.10256K} Kulkarni, S.~R.\ 2005, astro-ph/0510256 


\bibitem[Livio \& Riess(2003)]{2003ApJ...594L..93L} Livio, M., \& Riess, 
A.~G.\ 2003, ApJL, 594, L93 

\bibitem[Munari \& Zwitter(1997)]{1997A&A...318..269M} Munari, U., \& 
Zwitter, T.\ 1997, A\&A, 318, 269 



\bibitem[Ober et al.(1983)]{1983A&A...119...61O} Ober, W.~W., El Eid, 
M.~F., \& Fricke, K.~J.\ 1983, A\&A, 119, 61 

\bibitem[Oke et al.(1995)]{1995PASP..107..375O} Oke, J.~B., et al.\ 1995, 
PASP, 107, 375

\bibitem[Peng et al.(2002)]{2002AJ....124..266P} Peng, C.~Y., Ho, L.~C., 
Impey, C.~D., \& Rix, H.-W.\ 2002, AJ, 124, 266 

\bibitem[Pickles(1998)]{1998PASP..110..863P} Pickles, A.~J.\ 1998, PASP, 
110, 863 

\bibitem[Prieto et al.(2006)]{2006CBET..648....1P} Prieto, J.~L., 
Garnavich, P., Chronister, A., \& Connick, P.\ 2006, Central Bureau lectronic Telegrams, 648, 1 (2006).~ Edited by Green, D.~W.~E., 648, 1 

\bibitem[Quimby(2006)]{2006IAUC..8754....1Q} Quimby, R.\ 2006, Central 
Bureau Electronic Telegrams, 644, 1 (2006).~ Edited by Green, D.~W.~E., 
644, 1 

\bibitem[Rigon et al.(2003)]{2003MNRAS.340..191R} Rigon, L., et al.\ 2003, 
MNRAS, 340, 191 


\bibitem[Schlegel et al.(1998)]{1998ApJ...500..525S} Schlegel, D.~J., 
Finkbeiner, D.~P., \& Davis, M.\ 1998, ApJ, 500, 525 

\bibitem[Shelton(1997)]{1997SPIE.3126...455T} Shelton, J.~C. 1997,
SPIE, 3126, 455 

\bibitem[Skrutskie et al.(2006)]{2006AJ....131.1163S} Skrutskie, M.~F., et 
al.\ 2006, AJ, 131, 1163 

\bibitem[Smith et al.(2006)]{2006astro.ph.12617S} Smith, N., et al.\ 2006, astro-ph/0612617 


\bibitem[Taam \& Ricker(2006)]{2006astro.ph.11043T} Taam, R.~E., \& Ricker, 
P.~M.\ 2006, astro-ph/0611043 

\bibitem[Troy et al.(2000)]{2000SPIE.4007...31T} Troy, M., et al.\ 2000, 
SPIE, 4007, 31 

\bibitem[Tsujimoto \& Shigeyama(2006)]{2006ApJ...638L.109T} Tsujimoto, T., 
\& Shigeyama, T.\ 2006, ApJL, 638, L109 


\bibitem[Turatto et al.(2003)]{2003fthp.conf..200T} Turatto, M., Benetti, 
S., \& Cappellaro, E.\ 2003, From Twilight to Highlight: The Physics of 
Supernovae, 200 

\bibitem[Vazdekis(1999)]{1999ApJ...513..224V} Vazdekis, A.\ 1999, ApJ, 
513, 224 

\bibitem[Wegner et al.(2003)]{2003AJ....126.2268W} Wegner, G., et al.\ 
2003, AJ, 126, 2268 


\bibitem[Wood-Vasey \& Sokoloski(2006)]{2006ApJ...645L..53W} Wood-Vasey, 
W.~M., \& Sokoloski, J.~L.\ 2006, ApJL, 645, L53 

\bibitem[Woosley \& MacFadyen(1999)]{1999A&AS..138..499W} Woosley, S.~E., 
\& MacFadyen, A.~I.\ 1999, A\&AS, 138, 499 

\bibitem[Yoon \& Langer(2004)]{2004A&A...419..623Y} Yoon, S.-C., \& Langer, 
N.\ 2004, A\&A, 419, 623 

\end{thebibliography}
\end{document}